\documentstyle{article}
\begin{document}
\input{psfig}
\begin{center}
\Large\bf{Increase of Effective String Tension and Production of Strange
Particles}
 \\
\vspace{1cm}\large
  Tai An$^2$ and Sa Ben-Hao$^{1,3,4}$ \\
\begin{tabbing}
ttttt \= tt \= \kill
\>1. CCAST (World Lab.), P. O. Box 8730 Beijing, 100080 \\
\> \>China. \\
\>2. Institute of High Energy Physics, Academia Sinica, \\
\> \> P. O. Box 918, Beijing, 100039 China.\footnotemark \\       
\>3. China Institute of Atomic Energy, P. O. Box 275 (18), \\
\> \>Beijing, 102413 China.\\
\>4. Institute of Theoretical Physics, Academia Sinica, \\
\> \>Beijing 100080 China.\\

\end{tabbing}
\end{center}
\footnotetext{mailing address. \\Email: taian@hptc1.ihep.ac.cn}
\normalsize
\large
\begin{abstract}
The increase of effective string tension as a result of the hard gluon 
kinks on a string is investigated using a parametrization form. In this form 
the effective string tension increasing with energies in hadron-hadron 
collisions is due to the mini-jet (gluon) production in the collisions. 
The data of the energy dependence of the strange quark suppression factor 
in hh collisions are very well reproduced with this mechanism. Meanwhile, the 
experimental phenomena of approximate energy independence of the strange quark 
suppression factor in e$^+$e$^-$-annihilations are discussed.


\end{abstract}
\baselineskip 0.4cm
\parindent=0.3cm
\parskip=0.3cm
\hspace{0.3cm}
\newpage
Originally, in the LUND fragmentation scheme \cite{lund}, the strange quark 
suppression factor ($\lambda$ hereafter), i.e. the suppression of s quark pair 
production in the color field with respect to u or d pair production, was 
assumed to be a `constant'. This assumption was confirmed by e$^+$e$^-$ 
experiments up to Z$^0$ energy, except at low energies.  
\cite{kapa} \cite{opal}. 

However, it has been known for years that in hadron-hadron collisions 
$\lambda$ is not a constant but increasing from a value of 0.2 at the ISR 
energies to about 0.3 at the top of the SPS energies \cite{kapa}. Recent data 
of the p$\bar{\mbox{p}}$ collision at $\sqrt{s}$=630 GeV from UA1 again 
confirmed this fact \cite{ua1}. The $\lambda$ value 
extracted by UA1 is about 0.29. The energy dependence of $\lambda$ indicates 
that there is a mechanism in hh collisions which leads to enhanced production 
of strange quark pairs at higher energies. So far, no quantitative description 
exists for the issue of the energy dependence behaviours of $\lambda$ 
shown in the hh collisions and $e^+ e^-$-annihilations, due to our knowledge.
  
In this letter we propose a scenario to link the increase of the 
strange quark suppression factor to the effective string tension in the hh 
collisions based on the Lund string fragmentation model (more specifically, 
FRITIOF model \cite{fritiof} or its extension LUCIAE model \cite{luciae}). The 
idea of this scenario is that the production of mini-jets (gluons) in a 
hh collision will increase the effective string tension, therefore enhance the 
production of s quark pairs in the string fragmentation. A parametrization 
form is proposed based on such a scenario, which gives a reasonable energy 
dependence of the strange quark suppression factor in hh collisions. This 
scenario thus provides a dynamic explanation for the reduction of strange 
quark suppression at higher energies shown in the experimental data of hh 
collisions.
  
The string tension is defined as the energy per unit length of a string. 
However, the existence of gluons on a string (regarded as the transverse 
excitation or `kink' on a string in the Lund string fragmentation model) would 
wrinkle the string and give a fractal structure. Such a wrinkled string has 
obviously larger energy density in comparison with a string without gluon, 
thereby an enhanced string tension effectively \cite{torbjon}. 

The increase of the effective string tension with the appearance of gluons on 
a string might also be
understood in two following  ways: First, the string tension $\kappa$ 
is linked with the Regge slope $\alpha^\prime_R $ by $\kappa \sim 1/\alpha^
\prime_R$. When the pomeron exchange, which is generally looked upon as 
multigluon exchange, becomes important at higher energies the experiments  
extracted a smaller slope for the pomeron process i.e. $\alpha^\prime_P < \alpha
^\prime_R$. That means the pomeron process corresponds to the exchange of a 
string with larger string tension. Second, if we are allowed to compare QCD 
with an abelian theory the argument that the mean energy density in an 
abelian theory is only a function of ``the charge'' in the field would yield 
$\kappa \sim (Q_q^2 + 2Q_qQ_g)/A$, where $Q_q$ and $Q_g$ are the charges of 
quark field and gluon field, respectively and $A$ is the transverse size of a 
string. Above expression indicates again 
that the existence of the gluon field would enhance the string tension.
  
Since the question that we are concerned with is in the non-perturbative 
regime we can not establish a relation between the effective string tension 
and the gluons on a string from the first principle. Instead we will include 
this effect in a parametrization form.

In the Lund string fragmentation model the mean multiplicity from the 
fragmentation of a $q\bar{q}$ string state with mass $\sqrt{s}$ is given by
\begin{equation}
  \bar{n} \propto \ln (\frac{s}{s_{0}}),
\label{f1}
\end{equation}
where the parameter $\sqrt{s_{0}}$ is of the order of a typical hadron mass. 
However for a multigluon string state with (n-2) gluons, indexed in a colour 
connected way from the $q$ (index 1) to the $\bar{q}$ (index n), the 
corresponding mean multiplicity is 
 \begin{equation}
  \bar{n} \propto \ln (\frac{s}{s_{0}}) + \sum_{j=2}^{n-1} \ln (\frac{k_{\perp 
j}^2}{s_{0}}),
\label{f2}
\end{equation}
where $k_{\perp j}$, j=1,...(n-1), are the transverse momenta of the emitted 
gluons with $k_{\perp j}^2 \geq s_{0}$. The second term of Eq.(\ref{f2}) 
attributes to the contribution of gluons. Since the fractal structure of a 
string is governed by the hardest gluon on the string we then construct a 
quantity 
\begin{equation}
\xi =\frac{\ln(\frac{k_{\perp max}^2}{s_{0}})}{\ln (\frac{s}{s_{0}}) + 
\sum_{j=2}^{n-1} \ln (\frac{k_{\perp j}^2}{s_{0}})}
\label{f3}
\end{equation}
to represent the scale that the multigluon string is deviated from a pure 
$q\bar{q}$ string. We apply further the following expression 
to parametrize the effective string tension of the multigluon string
\begin{equation}
\kappa_{eff}=\kappa_{0} (1-\xi)^{-\alpha},
\label{f4}
\end{equation} 
where $\kappa_{0}$, the string tension of the pure $q\bar{q}$ string, is a 
constant and $\alpha$ is a parameter to be determined in comparison with data. 

In the Lund string fragmentation model, the $q\bar{q}$ pairs with the quark 
mass $m$ and the transverse momentum $p_{t}$ are produced from the colour 
field by a quantum tunneling process with probability
\begin{equation}
\exp(\frac{-\pi m^{2}}{\kappa_{eff}})\exp(\frac{-\pi p_{t}^{2}}{\kappa_{eff}}).
\label{f5}
\end{equation}
The above equation shows that the probability of the $s\bar{s}$ pair 
production with respect to a $u\bar{u}$ (or $d\bar{d}$) pair as well as the 
probability of a high $p_{t}$ $q\bar{q}$ pair production will be enhanced in a 
field with larger $\kappa_{eff}$. 

Assume that the width of the Gaussian transverse momentum distribution of 
$q\bar{q}$ pairs in the string fragmentation and the strange quark suppression 
factor of a string with effective string tension $\kappa_{eff1}$ are $\sigma_
{1}$ and $\lambda_{1}$, respectively, then those quantities of a string with 
effective string tension $\kappa_{eff2}$ can be calculated from Eq.(\ref{f5}), 
i.e.  
\begin{eqnarray}
\sigma_{2} &=& \sigma_{1}(\frac{\kappa_{eff2}}{\kappa_{eff1}})^{1/2}\nonumber\\
\lambda_{2} &=& \lambda_{1}^ {\frac{\kappa_{eff1}}{\kappa_{eff2}}}.
\label{f6}
\end{eqnarray}

We see here that $\sigma$ and $\lambda$ of above two string states are related 
by the ratio of the effective string tensions of those two string states only. 
It should be noted that the discussion above is also valid for the production 
of the diquark pairs from the string field, i.e. the production of the diquark 
pairs with respect to the $q\bar{q}$ pairs will be enhanced from a string
with larger $\kappa_{eff}$, therefore, more baryons (or antibaryons) will be 
formed in the final state. 

In FRITIOF a hh collision is pictured as the multi-scattering among the partons 
inside the two colliding hadrons. During the collision two hadrons are excited 
due to longitudinal momentum transfers and/or a QCD parton-parton scattering 
(Rutherford Parton Scattering, RPS) if four-momentum transfer is large. The 
highly excited states will emit bremsstrahlung gluons according to the colour-
dipole formalism until the $k_{\perp}$ of the emitted gluon is smaller than a 
given cut-off ($\sim 1$GeV/c). Thus a multigluon string state is formed and it 
is afterwards allowed to decay into final state hadrons according to the Lund 
fragmentation scheme implemented in JETSET generator \cite{jetset}. LUCIAE 
\cite{luciae} is developed based on FRITIOF by adding in more physical 
mechanisms like the rescattering of produced particles, the one in this paper, 
etc.. In $e^+e^- $ annihilation a multigluon string state is built up by the
standard QCD parton-shower scheme, which is also implemented in JETSET. 

In JETSET routine which runs together with LUCIAE event generator 
there are model parameters PARJ(2) (the same as $\lambda$) and PARJ(3). 
PARJ(3) is the extra suppression of strange diquark production compared to the 
normal suppression of strange quark pair. Both PARJ(2) and PARJ(3) are 
responsible for the s quark (diquark) suppression and related to the effective 
string tension. Besides $\lambda$ and PARJ(3) there is 
PARJ(1), which stands for the suppression of diquark-antidiquark pair 
production in the color field in comparison with the quark-antiquark pair 
production and is related to the effective string tension as well.  Another 
parameter PARJ(21) (the same as $\sigma$), which is the width of the Gaussian 
transverse momentum distribution of $q\bar{q}$ pairs in the string 
fragmentation, varies with $\kappa_{eff}$ too, but it is not related to the 
strange particle production directly.

It has been shown in the literature \cite{satai}\cite{topor}\cite{heinz} that 
the string fragmentation by JETSET with default values of PARJ(1)=0.1, PARJ(2)
=0.3 and PARJ(3)=0.4 overestimates the yield of strange particles in the pp
collision at 200 GeV/c. Thus in this letter we first retune these parameters
by comparing with the pp data of strange particle production \cite{hansen}. A 
new set of parameters PARJ(1)=0.046, PARJ(2)=0.2 and PARJ(3)=0.3 are found for 
pp at 200 GeV/c. We also give a new value of 0.32 for PARJ(21) (the 
corresponding default value is 0.37). In addition, in order to provide a value 
of $\lambda \simeq 0.3$ required by the p$\bar{\mbox{p}}$ data at the SPS 
energies \cite{kapa}, the parameter $\alpha$ in Eq.(\ref{f4}) and the 
$\sqrt{s_{0}}$ in Eq.(\ref{f3}) are determined to be about 3.5 and 0.8 GeV, 
respectively. Once we have determined the JETSET parameter values at 200 GeV/c 
we can then apply Eq.(\ref{f4}) and Eq.(\ref{f6}) to calculate the values of 
these parameters for other energies. 

Taking the changing behavior of those JETSET parameters, mentioned above, into 
account we have compared the calculated results of particle production with 
the data in Tab.1 and Tab.2 for the pp collision at 200 GeV/c 
\cite{hansen} and p$\bar{\mbox{p}}$ at $\sqrt{s}$=540 GeV \cite{540}, 
respectively. The agreement between the data and the results of LUCIAE is 
good. The four JETSET parameters for p$\bar{\mbox{p}}$ at $\sqrt{s}$=540 GeV 
are found to be PARJ(1)=0.115, PARJ(2)=0.304, PARJ(3)=0.406, PARJ(21)=0.376 
(the corresponding JETSET parameters for the pp collision at 200 GeV/c have  
already been determined in the last paragraph). The HIJING results \cite{topor} 
are also included in Tab.1 and Tab.2 for comparison. HIJING used JETSET with 
default parameters (i.e. $\lambda$ =0.3 etc.) to fragment a string, so it 
overestimates the production of strange particles in pp at 200 GeV/c, but 
reproduces the multiplicities of strange particles in p$\bar{\mbox{p}}$ at 
$\sqrt{s}$=540 GeV. 

The energy dependence of $\lambda$ in hh collisions is also calculated and 
the results are shown in Fig.1 together with the experimental data from 
\cite{kapa}. One sees from this figure that the agreement between data and 
the results of LUCIAE is reasonably good.

In hh collisions there are two strings formed before fragmentation. The 
$\lambda$ value calculated above is the mean value of the two strings. 
It is needed to point out that the increase of the effective string tension 
with increasing of energies in hh collisions is due to the production of high 
$k_{\perp}$-gluons from either the RPS or the bremsstrahlung radiation of the 
colour dipoles (both of which have been included in FRITIOF and/or LUCIAE). 

In Fig.2 we give the calculated results (solid line) of the energy dependence 
of $\lambda$ in $e^+e^-$ -annihilations and the corresponding data from \cite
{kapa} (black circle) and \cite{opal} (open circle). One sees from this figure 
that the constant $\lambda$ value around 0.3 only occurs in the energy region 
of $\sqrt{s} >$ 30 GeV and the agreement between data and calculated results 
is good. However, due to the large error bars of the data points in the low 
energy region it is hard to make any conclusion for the energy dependence of 
$\lambda$ in this region. But the threshold effect is seen 
obviously here below $\sqrt{s} < $ 30 GeV, while in hh collisions the 
saturation comes much later at $\sqrt{s} \simeq $ 200 GeV. 

When Lund model was developed to study hh collisions it was assumed that a 
string formed in hh collisions (quark-diquark string) is the same type of the 
string as formed in  $e^+e^-$- annihilations (quark-antiquark string) $-$ the 
jet universality. But detailed analysis reveals certain differences of this two 
types of string states due to the different physical processes involved in the 
formation of the strings. First, contrary to an $e^+ e^-$-annihilation 
the energy and momentum in a hh collision are not localised in a pointlike 
color charge but spread over some extended region. Any bremsstrahlung emission 
>from an extended source will necessarily be influenced by this extension 
(`the form factor'). Second, the hard RPS gluon jets produced in a hh 
collision will also greatly affect the topologic structure of a string formed. 
Third, only about half of the incoming energy can be used in the particle 
production in hadron-hadron collisions and the rest is carried away by the 
leading particles, unlike in an $e^+ e^-$-annihilation where all the incoming 
energy can be used to produce secondary hadrons. Those differences might 
explain the difference of the energy dependence behaviours of $\lambda$
in $e^+ e^-$-annihilations and hh collisions. 

It is worth pointing out that, since Eq.(\ref{f4}) can be approximated by 
$\kappa_{eff} \approx \kappa_{0}(1+\alpha\xi)$ thus in comparison with the 
formula from the abelian theory $\xi$ might be related to the ratio 
of charges of gluon and quark field $Q_{g}/Q_{q}$.

In summary, we have proposed a scenario inspired by the Lund string 
fragmentation model to investigate the reduction of strange quark 
suppression shown in experimental data of strange particle production in hh 
collisions. We obtain an appropriate parametrization form to 
characterize the relationship between the effective string tension and the 
hard gluons on a string. This scenario reproduces reasonably the energy 
dependence of the strange quark suppression factor in hh collisions and the 
nearly energy independence (above $\sqrt{s} >$ 30 GeV) in $e^+ e^-$-
annihilations. 

\begin{center}Acknowledgment\end{center}
This work is partly supported by the national Natural Science 
Foundation of China.

 \newpage 

\newpage
\begin{center}Figure Captions\end{center}
\begin{quotation}
Fig. 1 The energy dependence of $\lambda$ in hh collisions. The data 
are taken from \cite{kapa}.

Fig. 2 The energy dependence of $\lambda$ in  $e^+e^-$ -annihilations. The data points
are taken from \cite{kapa} (black circle) and \cite{opal}
(open circle), respectively.
\end{quotation}

\vspace{0.5cm}
\begin{tabular}{cccc}
\multicolumn{4}{c}{Table 1. Particle multiplicities (full phase space) for } \\
\multicolumn{4}{c} { the pp interaction at 200 GeV/c are compared with }\\
\multicolumn{4}{c}{experimental data \cite{hansen} and HIJING calculation.}\\
\hline\hline
 particle type & Expt. data & LUCIAE &HIJING \\
\hline
        $\pi^-$& 2.62$\pm$0.06  & 2.79&2.65\\
      $\pi^+$   &3.22$\pm$0.12& 3.24 & 3.23\\
      $\pi^0$   &3.34$\pm$0.24& 3.61  & 3.27\\
       $h^-$     &2.86$\pm$0.05& 3.01  & 3.03\\
        $K^+$     &0.28$\pm$0.06& 0.23  & 0.32 \\
         $K^-$       &0.18$\pm$0.05& 0.16  & 0.25\\
         $\Lambda + \Sigma^0$ &0.096$\pm$0.015&0.111 &0.165  \\
         $\overline\Lambda + \overline\Sigma^0$ &0.013$\pm$0.005&0.015& 0.037\\
         $K^0_{s}$ &0.17$\pm$0.01&0.18 & 0.27\\
         $p$ &1.34$\pm$0.15&1.00& 1.45\\
        $\overline p$ &0.05$\pm$0.02&0.04& 0.12\\
\hline\hline
\end{tabular}

\vspace{0.5cm}
\begin{tabular}{cccc}
\multicolumn{4}{c}{Table 2. Particle multiplicities (full phase space) for }\\
\multicolumn{4}{c}{the p$\bar{\mbox{p}}$ interaction at $\sqrt{s}$=540 GeV are compared}\\
\multicolumn{4}{c}{ with experimental data \cite{540} and HIJING calculation.}\\
\hline\hline
 particle type & Expt. data & LUCIAE &HIJING \\
\hline
        All charged & 29.4$\pm$0.3  &28.4 &28.2\\
      $K^0 + \overline K^0$   &2.24$\pm$0.16& 2.62  & 1.98\\
      $K^+ + K^-$   &2.24$\pm$0.16& 2.72 & 2.06\\
       $p + \overline p$     &1.45$\pm$0.15& 2.62  & 1.55\\
        $\Lambda + \overline\Lambda $     &0.53$\pm$0.11& 0.58  & 0.50 \\
         $ \Sigma^+ + \Sigma^- + \overline\Sigma^+ + \overline\Sigma^-$ &0.27$\pm$0.06& 0.32  & 0.23\\
         $\Xi^-$ &0.04$\pm$0.01&0.032 &0.037\\
         $\pi^+ + \pi^-$ &23.9$\pm$0.4&22.69& 23.29\\
         $<K^0_{s}>$ &1.1$\pm$0.1&1.32 & 0.99\\
         $\pi^0$ &11.0$\pm$0.4&12.82& 13.36\\
\hline\hline
\end{tabular}


  \hbox{
    \vbox{

}
  }



  \hbox{
    \vbox{

}
  }


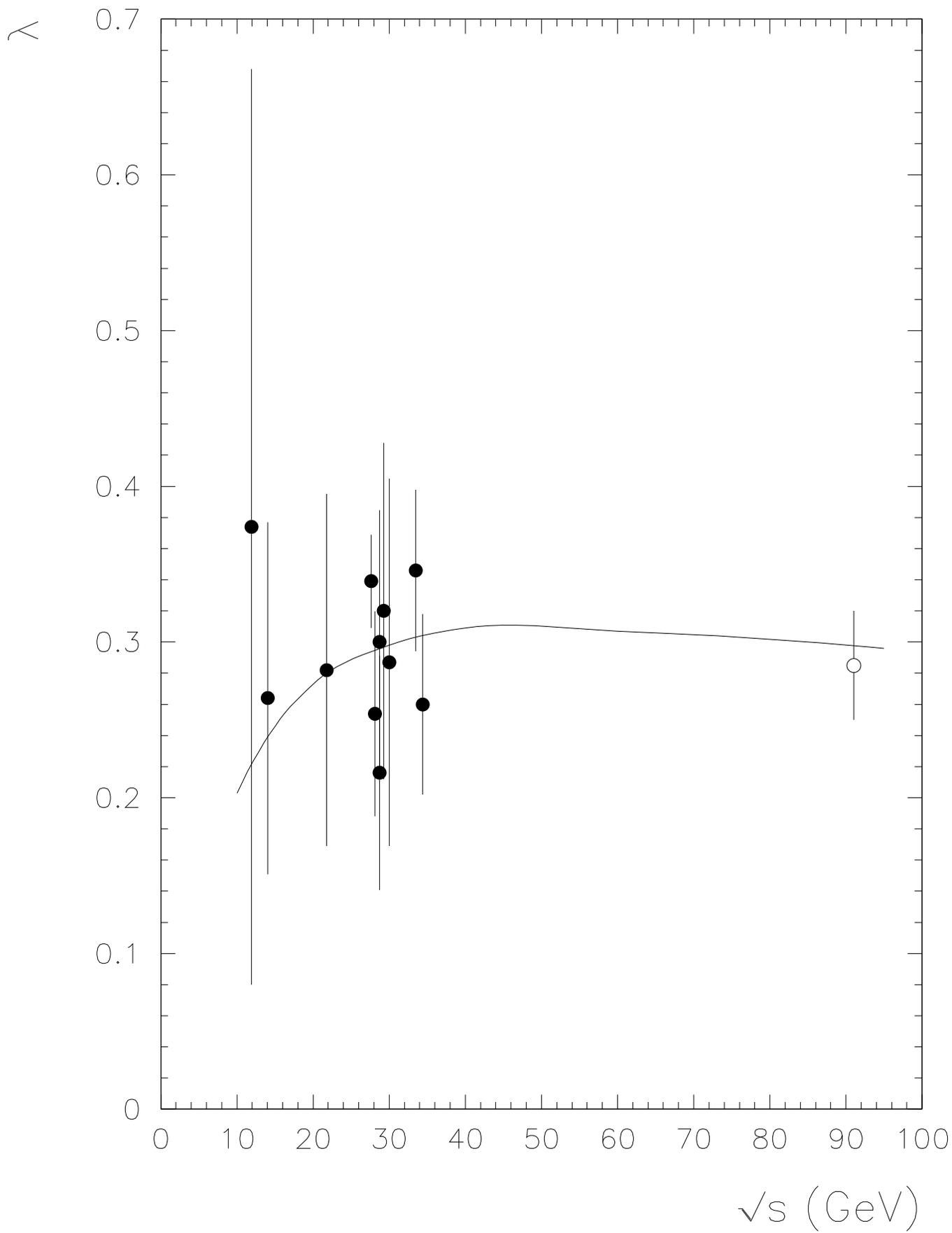
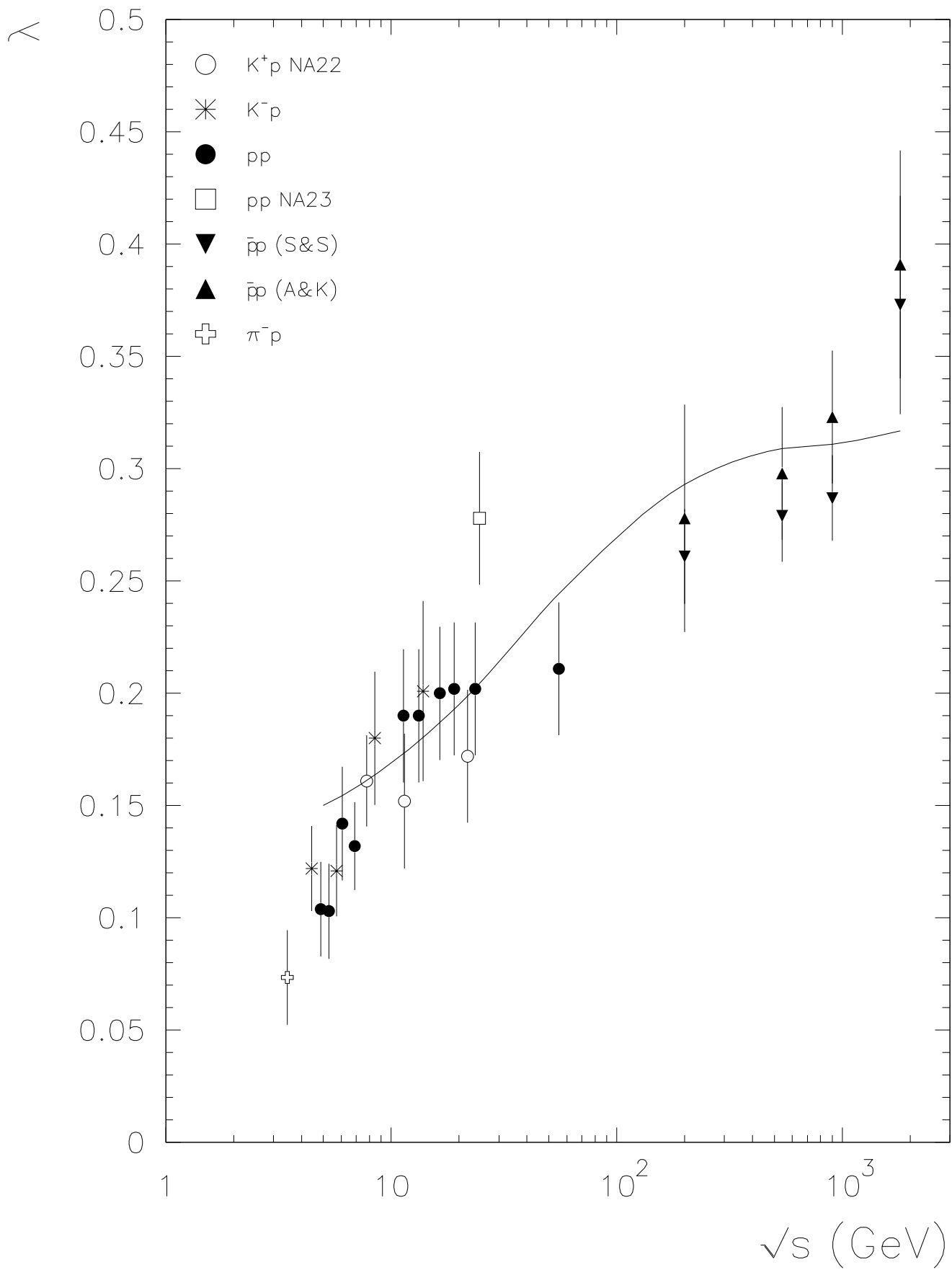
\end{document}